\documentclass[3p,times,twocolumn]{elsarticle}

\usepackage{ecrc}

\volume{00}

\firstpage{1}

\journalname{Nuclear and Particle Physics Proceedings}

\runauth{}

\jid{nppp}

\jnltitlelogo{Nuclear and Particle Physics Proceedings}

\usepackage{amssymb}

\usepackage[figuresright]{rotating}

\begin{document}

\begin{frontmatter}



\dochead{}

\title{Direct photon yield in pp and in Pb-Pb collisions measured with the ALICE experiment}


\author{Davide Francesco Lodato (for the ALICE collaboration)}

\address{Utrecht University, Princetonplein 5, Utrecht, the Netherlands}
\address[E-mail:]{davide.francesco.lodato@cern.ch}
\begin{abstract}
The measurement of direct photon production in Pb-Pb collisions at $ \sqrt{s_{\rm{NN}}} = 2.76 $ TeV and in pp collisions at $ \sqrt{s}= 7 $ TeV with the ALICE experiment is presented. In Pb-Pb collisions a clear direct photon signal below 3 GeV/$c$ is observed only for the $ 0-20 \mbox{ }\%$ most central collisions. No excess is observed for semi-central and peripheral Pb-Pb collisions and in pp collisions.\\ 
Furthermore, in pp collision the measurement of direct photon production in the range 10 $\leq p_{\rm{T}} \leq$ 60 GeV/$c$ is reported. The analysis is performed on the EMCal-triggered data taken in $2011$. The two main sources of background, namely photons from fragmentation processes and decay photons, have been subtracted from the inclusive photon spectrum by means of an isolation technique combined with the study of the transverse dispersion of electromagnetic showers. The measurement is in agreement both with NLO pQCD calculations and with those performed by the ATLAS 
and the CMS collaborations, 
extending the isolated photon spectrum investigated at LHC towards lower values of $p_{\rm{T}}$.
\end{abstract}


\end{frontmatter}



\section{Introduction}
\label{sec:Intro}
In pp collisions the measurement of direct photon production can be used to test both the pQCD calculations and the binary scaling behaviour of the initial hard scattering. At high $p_{\rm{T}}$, pQCD processes like quark-gluon Compton scattering and quark-antiquark annihilation are the main contribution to direct photon production, and allow us to probe directly the gluons within hadrons. However, photons are also produced in jet fragmentation processes, in which part of the information about the hard scattering dynamics is lost. Reducing the contribution of photons from the latter source via isolation techniques helps to better constrain the gluon parton distribution function \cite{Helenius:2014qla}.

In nucleus-nucleus collisions direct photons are produced at every stage of the collision and therefore are sensitive to the different phases of the medium evolution. The low-$p_{\rm{T}}$ component of the direct photon spectrum is dominated by thermal production in the quark-gluon plasma and in the hadron-gas phase, giving us access to information on the temperature of the hot and dense medium in which direct photons are produced.

For $p_{\rm{T}}$ greater than 5 GeV/$c$, direct photons are mainly produced in hard partonic scattering processes in the early stage of the collision, leaving the strongly interacting medium unscathed and provide access to information about the initial dynamics.\\



\section{Low-$p_{\rm{T}}$ direct photon measurement in pp and Pb-Pb collisions: $R_{\gamma}$}
\label{sec:LowpT}
The inclusive photon yield is measured both directly via the calorimetric method with the ALICE PHOton Spectrometer (PHOS) and via the reconstruction of photons via Photon Conversion Method (PCM). In the latter case a secondary vertex finder is used to pair electron positron tracks with a large impact parameter. Several selection criteria, like constraints on the opening angle and on the reconstructed invariant mass, are applied in order to optimize the signal to background ratio. PHOS is a  highly granulated leadt ungstate (PbWO$_{4}$) homogeneous calorimeter with a coverage of $\Delta\varphi = 60^{\circ}$ and $|\eta|< 0.12$. The analysis performed via PCM makes use of the $(\eta-\varphi)$ coverage of the ALICE Time Projection Chamber \cite{Abelev:2014ffa}, respectively $|\Delta\eta| < 0.9$ and full azimuthal coverage . 

The same strategy is used to analyse the data taken both in Pb-Pb at $\sqrt{s_{\rm{NN}}} = 2.76$ TeV and pp collisions at $\sqrt{s_{\rm{NN}}} = 7$ TeV. The Pb-Pb analysis is performed in three bins of centrality (central: $0-20\mbox{ }\%$, semi-central: $20-40\mbox{ }\%$ and peripheral: $40-80 \mbox{ }\%$). The direct photon signal is obtained by subtracting the contribution of decay photons from the inclusive photon spectrum: 
\begin{equation}
\gamma_{\rm{direct}} =\gamma_{\rm{inc}} - \gamma_{\rm{decay}} = \left( 1 - \frac{\gamma_{\rm{decay}}}{\gamma_{\rm{inc}}}\right) \cdot \gamma_{\rm{inc}},
\end{equation}
where the main sources of decay photons are neutral meson (mainly $\pi^0$ and $\eta$ via their 2-photon decay channel). 
The raw photon spectrum is corrected for purity, reconstruction efficiency and, in case of the PCM analysis, for the conversion probability in the detector material. 

In pp collisions the decay photon spectrum is calculated from a parametrization of the measured yield of $\pi^0$ and $\eta$ mesons, reconstructed via the 2-photon decay channel; in Pb-Pb the parametrization is available only for $\pi^0$. In both cases, additional sources of decay photons yields from other mesons like $\eta(')$, $\omega$, $\phi$ and $\rho_{0}$ are computed via $m_{\rm{T}}$-scaling and included into the cocktail calculations.

The decay photon contribution was subtracted by calculating the so-called double ratio $R_{\gamma} = (\gamma_{\rm{inc}}/ \pi^{0}_{\rm{param}})/(\gamma_{\rm{decay}}/\pi^{0}_{\rm{param}})$. By doing so, some uncertainties of the measurement cancel exactly. In pp collisions no direct photon signal is found \cite{Wilde:2012wc}. 

\begin{figure}[!h]
\includegraphics[width=0.44\textwidth]{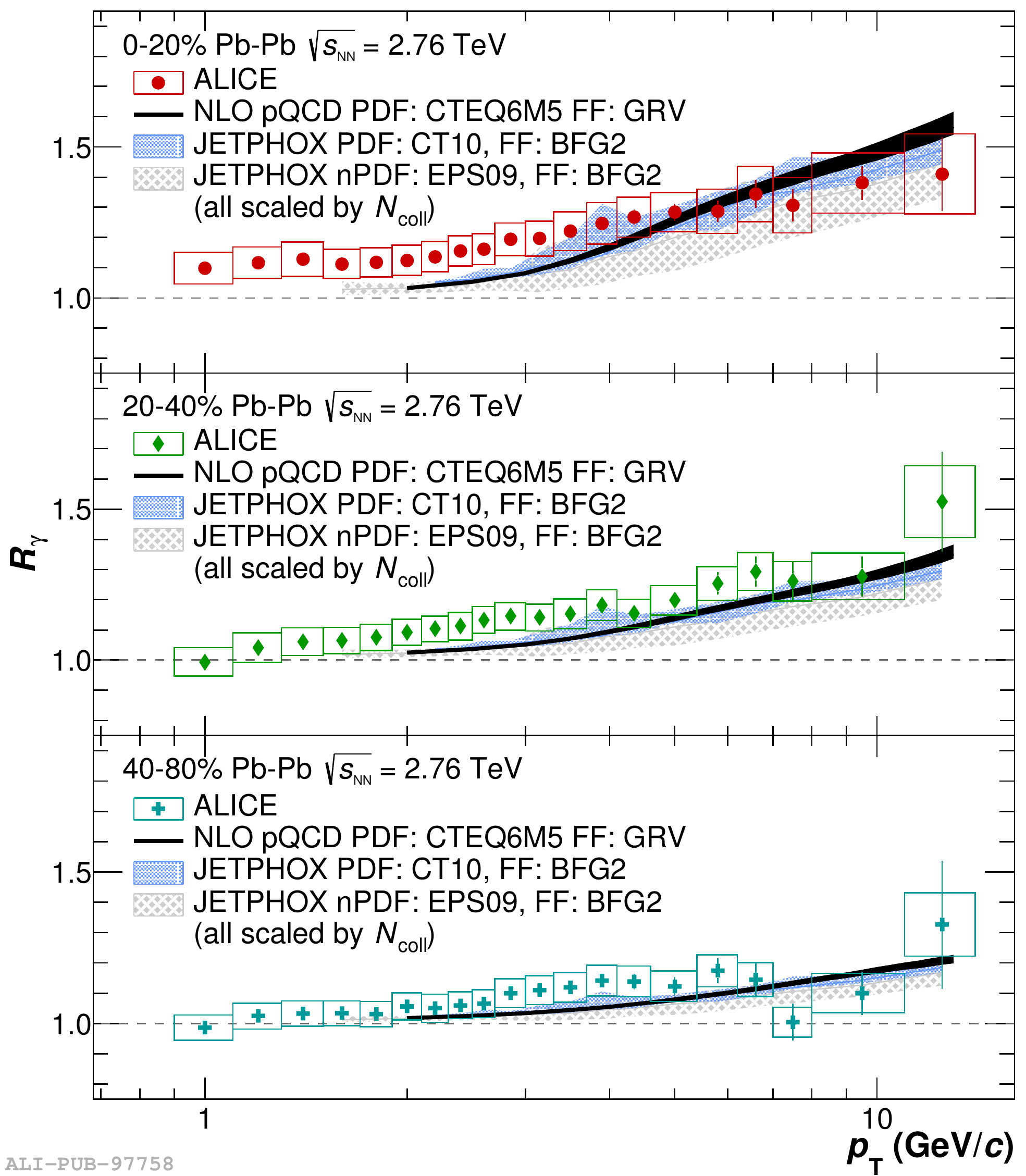}
\caption{Combined PCM and PHOS double ratio $R_{\gamma}$ for three different centrality classes. It is clear the increase of direct photon signal with the centrality of the collision, for $p_{\rm{T}} < 3 $ GeV/$c$.}
\label{fig:RgammaPbPb}
\end{figure}

\begin{figure}[!h]
\includegraphics[width=0.45\textwidth]{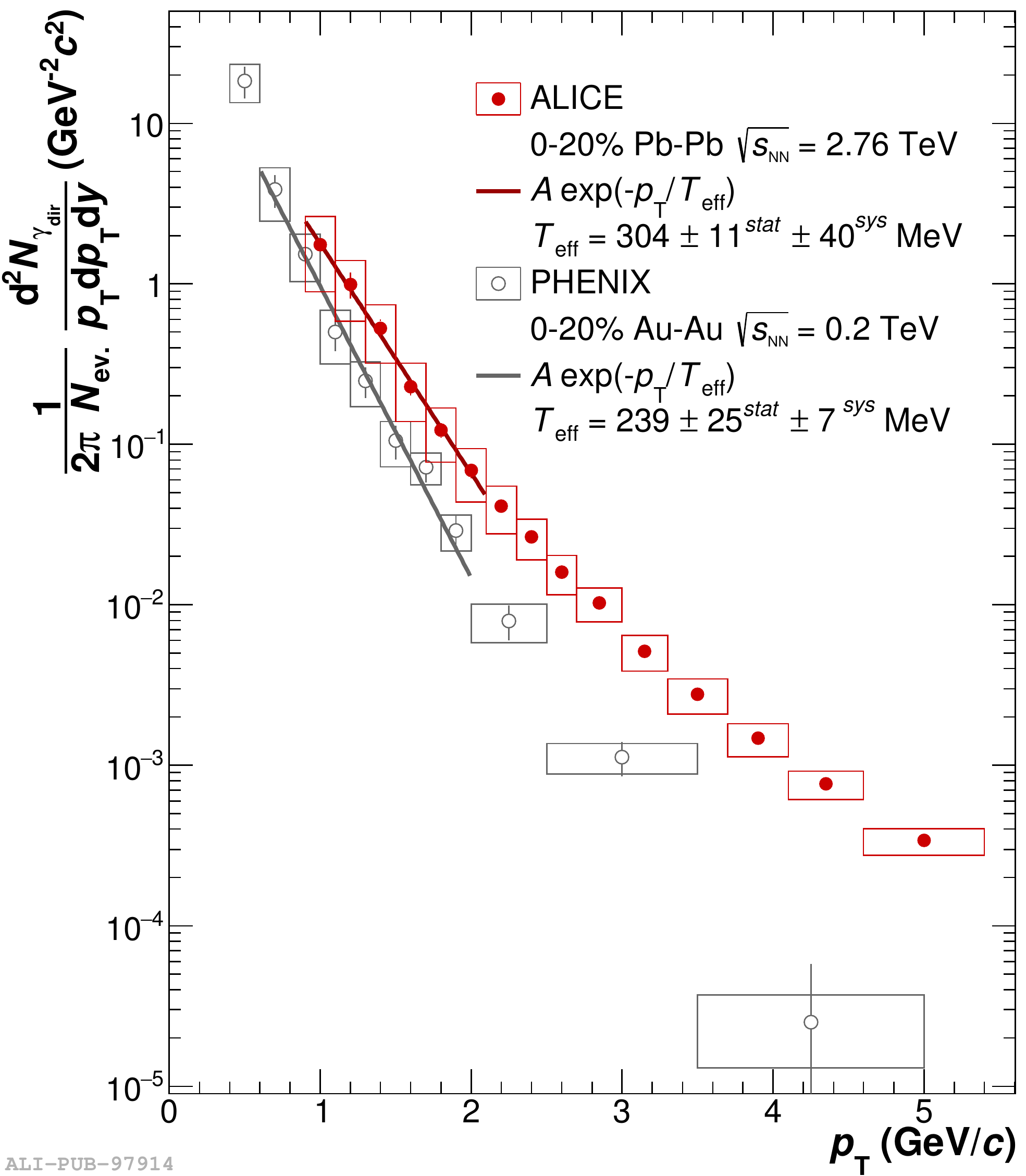}
\caption{Direct photon spectrum computed for the most central collisions and fitted to an exponential function. The effective temperature extracted from fitting the ALICE result is $T_{\rm{eff}} = 304 \pm 11^{\rm{stat}} \pm 40 ^{\rm{syst}}$ MeV.}
\label{fig:020Spec}
\end{figure}

The results of the measurements of $R_{\gamma}$ in Pb-Pb collisions are presented in Fig. \ref{fig:RgammaPbPb} for different centralities. For comparison, the NLO pQCD calculations for pp collisions and scaled by $N_{\rm{coll}}$ are overlayed. The results obtained for non-central collisions are found to be in agreement with the theoretical predictions; the direct photon signal increases with the centrality of the collision.
The shown $R_{\gamma}$ is obtained by combining the PHOS and PCM measurements, for which the two analysis reach an agreement of 0.4 standard deviations. More details on the analyses and their comparison and combination can be found in \cite{Adam:2015lda}. In Fig. \ref{fig:020Spec}, the direct photon $p_{\rm{T}}$ spectrum computed for the most central collisions is presented. In the range $1 < p_{\rm{T}} < 2$ GeV/$c$, a fit with an exponential function is performed to extract the effective temperature of the medium, $T_{\rm{eff}}$ from the inverse slope parameter, which is estimated to be $T_{\rm{eff}} = 304 \pm 11^{\rm{stat}} \pm 40 ^{\rm{syst}}$ MeV. 

As a comparison, results from PHENIX for the same measurement at $\sqrt{s_{\rm{NN}}}= 0.2$ TeV are also shown.

\section{Direct photon at high $p_{\rm{T}}$: Isolation technique for contamination estimate}
\label{sec:HighpT}
The analysis of direct photon production at high $p_{\rm{T}}$ has been carried out in pp collisions at $\sqrt{s}= 7 $ TeV by analysing the EMCal L0-triggered data collected by the ALICE experiment in 2011. The photons are measured directly via the calorimetric method with EMCal \cite{Abeysekara:2010ze} while charged particles are reconstructed by the full ALICE tracking system. The EMCal is a Pb-scintillator sampling calorimeter with a granularity of $\Delta\eta \times \Delta\varphi = 0.0143\times0.0143$ and a total coverage of $\Delta\varphi = 100^{\circ}$ and $|\eta|< 0.7$. 

The EMCal-L0 trigger correlates the energy deposited in the V0 detectors, placed at $2.8 < \eta < 5.1$ and $ -3.7 < \eta < -1.7$ on the A and C side respectively, with the energy measured by the EMCal detector and it is used to select events with a large energy deposition ($> 5 $ GeV)  in a $4\times4$ EMCal towers patch. This allows for a reduction of the data volume and of the detector dead time, enhancing, as a consequence, the integrated luminosity recorded.\\
Particles will deposit their energy in multiple towers (or cells) of the calorimeter. A clustering algorithm identifies cells (called seeds) with energy deposit $E_{seed} \geq $300 MeV, aggregating its neighbouring cells as long as $E_{cell} \geq$ 100 MeV. In order to select direct photon candidates, various selection criteria are applied. Clusters produced by charged particles are rejected on the base of a track-proximity criterion. \\
The production of direct prompt photons at high $p_{\rm{T}}$ is mainly due to quark-gluon Compton scattering and quark-antiquark annihilation. Neutral pions  represent the main source of background via the 2-photon decay process. In the range $10 < p_{\rm{T}} < 60$ GeV/$c$, the two photons will likely be reconstructed as single elongated cluster. Direct photon cluster selection is based on the study of the width of the energy distribution 
\begin{equation}
\hspace{-20pt} 
\sigma^{2}_{\rm{long}} = 0.5 \times (d_{\eta\eta} + d_{\varphi\varphi}) + \sqrt{0.25 \times(d_{\eta\eta} - d_{\varphi\varphi})^2 + d_{\eta\varphi}^2},
\end{equation} 
along the major axis of the reconstructed cluster. The dispersions $d_{ij}$ are computed with a logarithmic weighting as in \cite{Awes:1992yp}. 

\begin{figure}[!ht]
\includegraphics[width=0.5\textwidth]{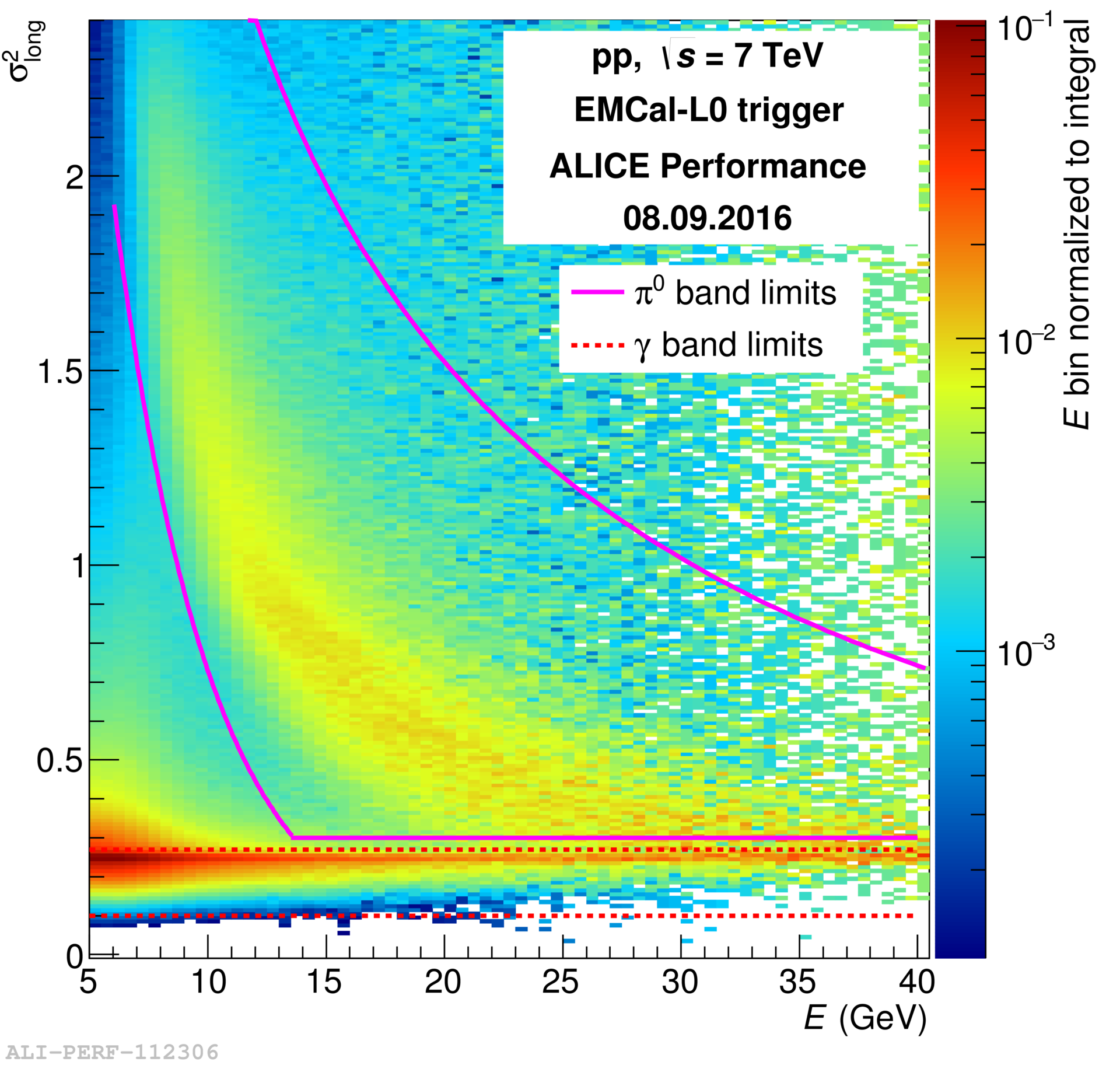}
\caption{$\sigma^{2}_{\rm{long}}$ distribution as a function of the (neutral) cluster energy. The magenta lines indicate the band containing merged photon clusters from $\pi^0$ decays and the red dashed lines indicate the single-photon band.}
\label{fig:M02}
\end{figure}

In Figure \ref{fig:M02} the dependence of the distribution of the $\sigma^{2}_{\rm{long}}$ parameter on the energy of the cluster is presented. Two main regions can be identified: single photon clusters dominate at $\sigma^{2}_{\rm{long}}  \leq 0.3$ while clusters from $\pi^0$ decay photons are clearly visible for $E \leq 20$ GeV/$c$. 

Isolation techniques can help to distinguish direct photons both from those produced in fragmentation processes and further reduces the decay photon background. The advantage of applying an isolation criterion is discussed in \cite{Ichou:2010wc}. The isolation criterion used in this analysis is based on the measurement of the total activity in a cone of radius $R=0.4$ around the selected photon candidates; the condition for which a cluster is considered isolated is:  $$E_{\rm{T}}^{\rm{cone}} = \Sigma(E_{\rm{T}}^{\rm{clust}} + p_{\rm{T}}^{\rm{tracks}}) < 2 \mbox{ GeV/}c.$$

Only clusters whose cone is fully contained within the EMCal acceptance are selected, reducing the $\eta-\varphi$ acceptance of the analysis to $|\eta|<0.27$ and $1.8 <\varphi < 2.7$. By making use of both the $\sigma^{2}_{\rm{long}}$ and the $E_{\rm{T}}^{\rm{cone}}$ distributions it is possible to estimate the contamination in our sample of isolated photon-like clusters by means of a double sideband method. The validity of this method is related to the fact that the $E_{\rm{T}}^{\rm{cone}}$ distribution is independent of $\sigma^{2}_{\rm{long}}$, for the background. Studies performed on MonteCarlo samples have shown the presence of cross-talk  between the $\sigma^{2}_{\rm{long}}$ and $E_{\rm{T}}^{\rm{cone}}$ variables. We can compute then a factor $\alpha_{MC}(p_{\rm{T}})$ and use it to correct the estimated contamination in data, which is then subtracted from the raw isolated photon spectrum. The remaining spectrum is corrected for reconstruction and identification efficiency. The differential cross section density is computed by scaling the corrected spectrum for trigger efficiency and total inelastic cross section measured by the ALICE collaboration~\cite{Abelev:2012sea}. The analysis has been repeated by varying different selection criteria in order to estimate the uncertainty related to various choices of cuts in the analysis. The main source of uncertainties is due to discrepancies in the modelling of the transverse shower shape distribution in MC with respect to data. 

\begin{figure}[!h]
\includegraphics[width=0.48\textwidth]{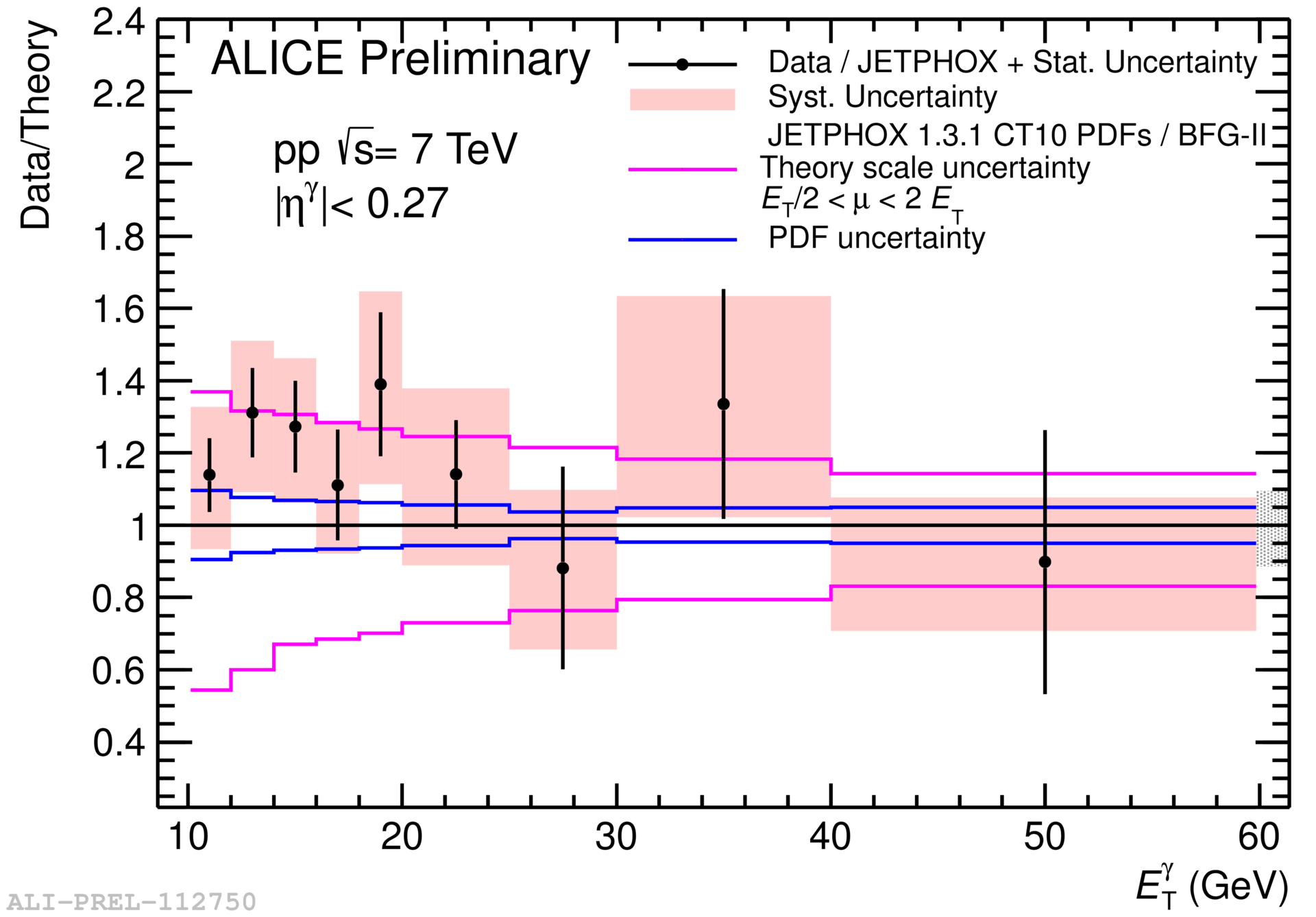}
\caption{Comparison between the isolated direct photon cross section measured by the ALICE collaboration and a NLO pQCD calculation. The agreement of the measured cross section is found to be reasonable in the whole investigated $p_{\rm{T}}$ range.}
\label{fig:ratioJETPHOX}
\end{figure}

Fig. \ref{fig:ratioJETPHOX} shows the comparison between the measured cross section and the theoretical calculation from JETPHOX \cite{Aurenche:2006vj}; a reasonable agreement is found in the whole $p_{\rm{T}}$ range investigated. This result extends the range investigated by the ATLAS  \cite{Aad:2010sp} and the CMS collaborations \cite{Chatrchyan:2011ue} down to $10 \mbox{ GeV/}c$ .

\section{Summary}
\label{sec:Summary}
In these proceedings we presented a summary of the direct photon measurements in pp and Pb-Pb collisions performed by the ALICE collaboration. 

In pp collision the direct photon yield is measured in the $p_{\rm{T}}$ range $[10-60] \mbox{ GeV/}c$ using the calorimetric method (EMCal) complemented by means of isolation techniques. All results obtained in pp collisions are in good agreement with NLO pQCD calculations. For lower values of the transverse momentum both the calorimetric method (PHOS) and the Photon Conversion Method are used. The signal has been studied by means of the double ratio $R_{\gamma}$~\cite{Wilde:2012wc}, and no direct photon signal has been found at low $p_{\rm{T}}$ in pp collisions.
 
In Pb-Pb collisions, the same low $p_{\rm{T}}$ part (up to 14 GeV/$c$) of the direct photon spectrum has been investigated. The measurement has been carried out with PHOS and PCM using the same analysis strategy applied to pp collisions. The Pb-Pb analysis is performed in three centrality classes. An excess in the $R_{\gamma}$ is found only for the most central ($0-20\mbox{ }\%$) collisions. A fit of the transverse momentum spectrum with an exponential function gives an inverse slope from which the effective temperature of the medium is estimated to be $T_{\rm{eff}} = 304 \pm 11^{\rm{stat}} \pm 40 ^{\rm{syst}}$.



\nocite{*}
\bibliographystyle{elsarticle-num}
\bibliography{LodatoD.bib}






\end{document}